\documentclass[prd,aps,nofootinbib]{revtex4}

\usepackage{graphicx}

\begin{document}

\title{On some hybrid-types of $Q$ balls in the gauge-mediated supersymmetry breaking}

\author{Shinta Kasuya$^a$ and Masahiro Kawasaki$^{b,c}$}

\affiliation{
$^a$ Department of Information Sciences,
     Kanagawa University, Kanagawa 259-1293, Japan\\
$^b$ Institute for Cosmic Ray Research,
     University of Tokyo, Chiba 277-8582, Japan\\
$^c$ Institute for the Physics and Mathematics of the Universe, 
     University of Tokyo, Chiba 277-8582, Japan}

\date{October 5, 2009}

\begin{abstract}
We revisit the new-type of the $Q$ ball (the gravity-mediation type of the $Q$ ball in the 
gauge-mediation), in order to clarify its properties and correct some misunderstandings 
found in the recent literature. In addition, we investigate the feature of the other kind of the 
hybrid-type of the $Q$ ball, which was considered in the context of the $Q$-ball capture 
by the neutron star.
\end{abstract}


\maketitle


\section{Introduction}
A $Q$ ball is a nontopological soliton, the minimum energy configuration of the (complex) 
scalar field, whose existence is guaranteed by non-zero charge $Q$ \cite{Coleman}. 
It appears naturally in supersymmetric theories \cite{KuSh,EnMc}. In particular in the 
gauge-mediated supersymmetry (SUSY) breaking, the $Q$ ball is stable against the decay 
into fermions and other scalars for large enough charge $Q$, and can be  
the dark matter of the universe \cite{KuSh}.
Such large $Q$ balls are naturally produced in the early universe as byproducts of the
Affleck-Dine mechanism for baryogenesis \cite{KuSh,KK1,KK2,KK3}. They could be detectable
and/or constrained by various experiments \cite{KKST,Arafune,KK3} and by considering 
astrophysically such as the capture by neutron stars \cite{NS1,NS2}.

The properties of the $Q$ ball is determined by the shape of the scalar potential. In the
gauge-mediated SUSY breaking, the potential is flat beyond the messenger scale, and
the mass grows as $M \propto Q^{3/4}$ \cite{Dvali}. As the field amplitude becomes large,
the potential will be dominated by the effect of gravity-mediation, and the features of the
$Q$ ball change such as $M\propto Q$, for example. We called this kind of the $Q$ ball 
the new-type $Q$ ball \cite{New}. As we mentioned in \cite{New}, the `metamorphosis' of the
$Q$ ball should take place smoothly. 

In this article, we revisit the properties of the new-type $Q$ balls, with special attention to 
clarify the transition region between the gauge and gravity mediation. This is partly because
we must correct some misunderstandings found in \cite{Shoemaker}, where they claim the 
new-type $Q$ ball disintegrates into the gauge-type ones. 

In addition, we also investigate the features of the other kind of hybrid type of the $Q$ ball 
considered in \cite{NS2}, where the gauge-type $Q$ ball changes to the thin-wall-type $Q$ 
ball\footnote{
In Ref.~\cite{NS2}, it is called the `curved direction' $Q$ ball. In this article, we call it the `thin-wall-type'
$Q$ ball because of its profile as shown in Fig.~\ref{gaugeNR-prof}.}
as it fattens in the interior of the neutron star. This happens since the field value inside
the $Q$ ball cannot grow further when the potential is lifted by nonrenormalizable operators 
at large field values.\footnote{This thin-wall-type $Q$ ball is not created through the 
Affleck-Dine mechanism, since the scalar field does not feel spatial instabilities while
it stays on the nonrenormalizable potential.}

\section{$Q$ ball solution}
Let us first review the general properties of the $Q$ ball in the scalar theory wtih
a global $U(1)$ symmetry. In the context of SUSY $Q$ balls, the charge is usually 
the baryon and/or lepton numbers. The energy and charge are given respectively by
\begin{eqnarray}
E & = & \int d^3x \left[ \partial_\mu \phi \partial^\mu \phi^* +V(\phi)\right], \\
Q & = & \frac{1}{i} \int d^3x (\dot{\phi}\phi^* - \phi \dot{\phi}^*),
\end{eqnarray}
where $V(\phi)$ is the potential. Since the $Q$ ball is the energy minimum configuration
of the scalar field with finite charge $Q$, using a lagrange multiplier $\omega$, we can write 
the energy as \cite{small}
\begin{equation} 
{\cal E}_\omega = E
+ \omega \left[ Q - \frac{1}{i} \int d^3x (\dot{\phi}\phi^* - \phi \dot{\phi}^*) \right].
\end{equation}
Energy minimum configuration is obtained when the solution is spherical symmetric and
rotating: $\phi(x) = \varphi(r) e^{i\omega t}/\sqrt{2}$. Then, the energy is rewritten as
\begin{equation} 
{\cal E}_\omega = 4\pi \int_0^\infty dr r^2\left[ \frac{1}{2}\left(\frac{d\varphi}{dr}\right)^2
+V(\varphi)-\frac{1}{2}\omega^2\varphi^2\right] + \omega Q.
\end{equation}
In order to obtain the energy minimum solution, we just have to solve the equation
\begin{equation}
\label{eom}
\frac{d^2\varphi}{dr^2}+\frac{2}{r}\frac{d\varphi}{dr}
+\left[\omega^2\varphi-\frac{dV}{d\varphi}\right]=0,
\end{equation}
 with boundary conditions $\varphi(\infty)=0$ and $\varphi'(0)=0$.
 
In the next sections, we apply the above argument and solve the equation (\ref{eom}) 
numerically for two kinds of hybrid-type $Q$ balls: One is the gauge-type and new-type
$Q$ balls, and the other is the gauge-type and thin-wall-type $Q$ balls.

\section{new-type $Q$ balls}
\begin{figure*}[h]
\begin{tabular}{cc}
\includegraphics[width=85mm]{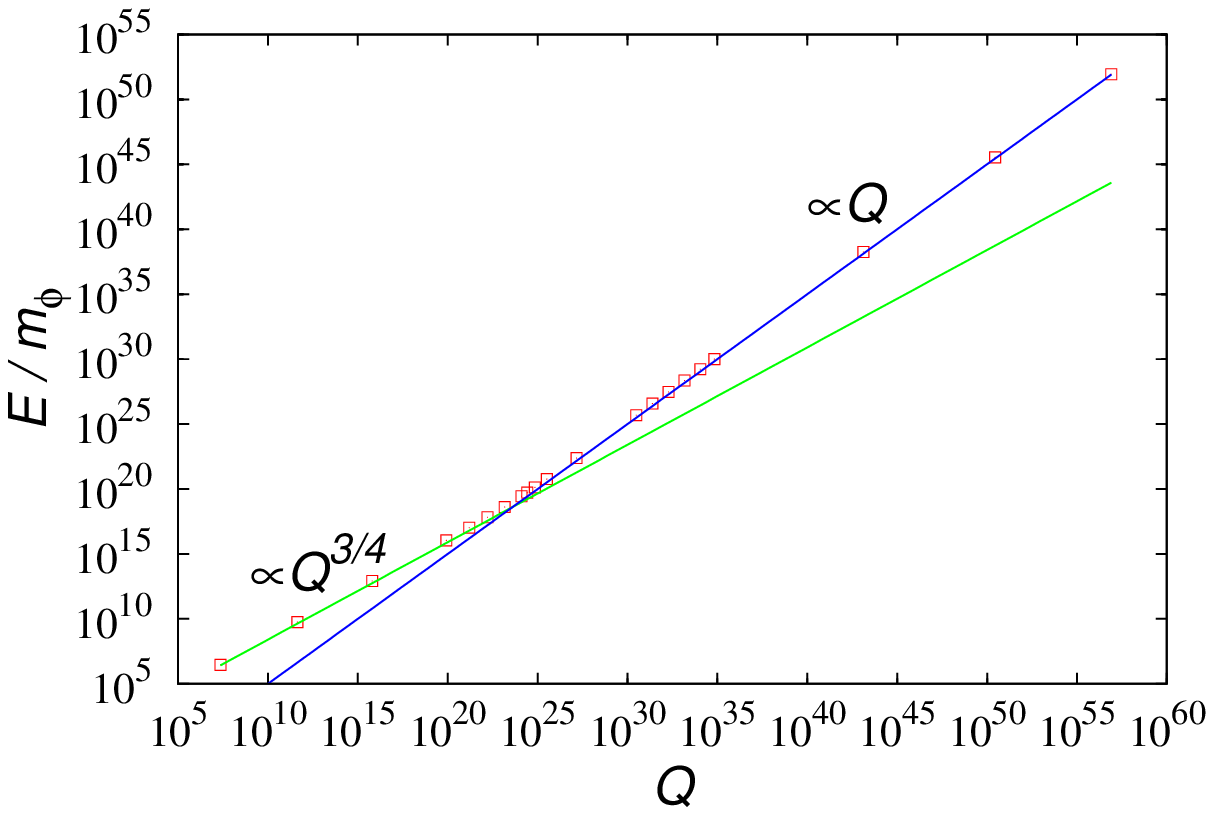} &
\includegraphics[width=85mm]{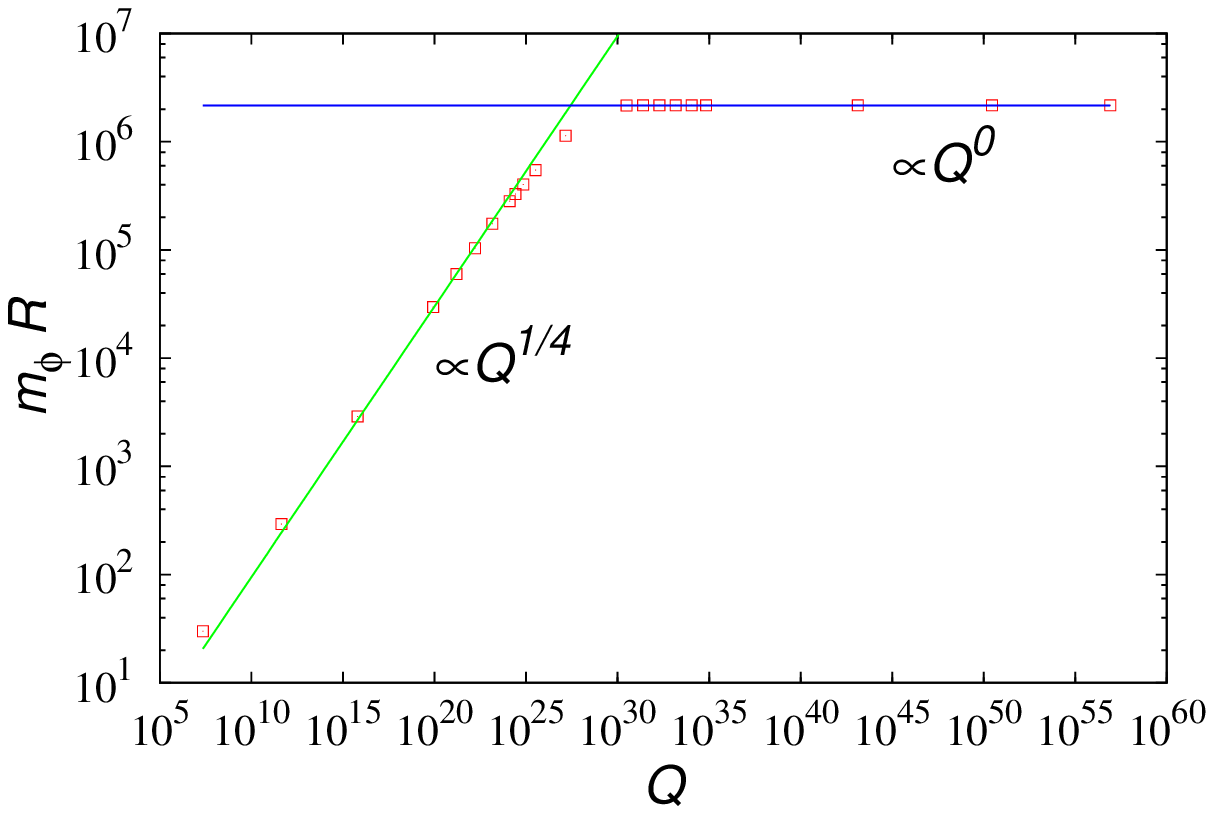} \\
\includegraphics[width=85mm]{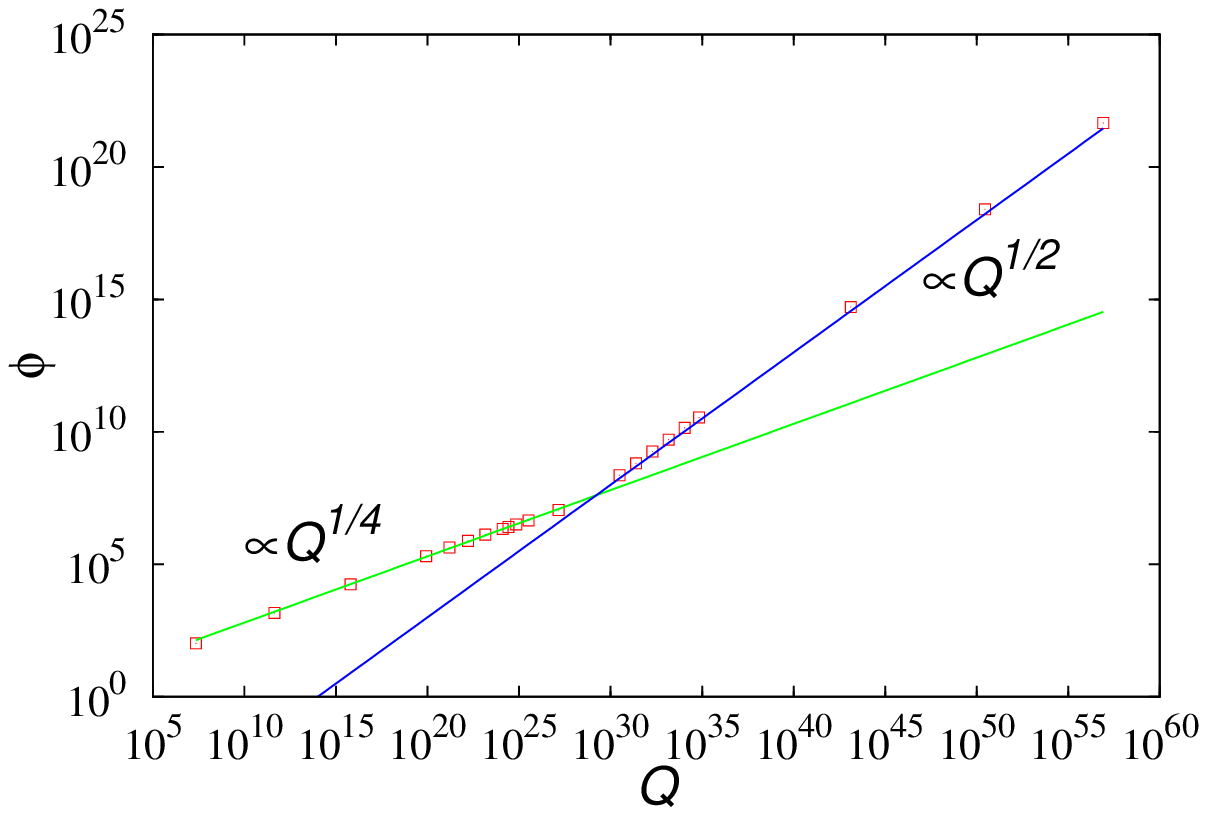} &
\includegraphics[width=85mm]{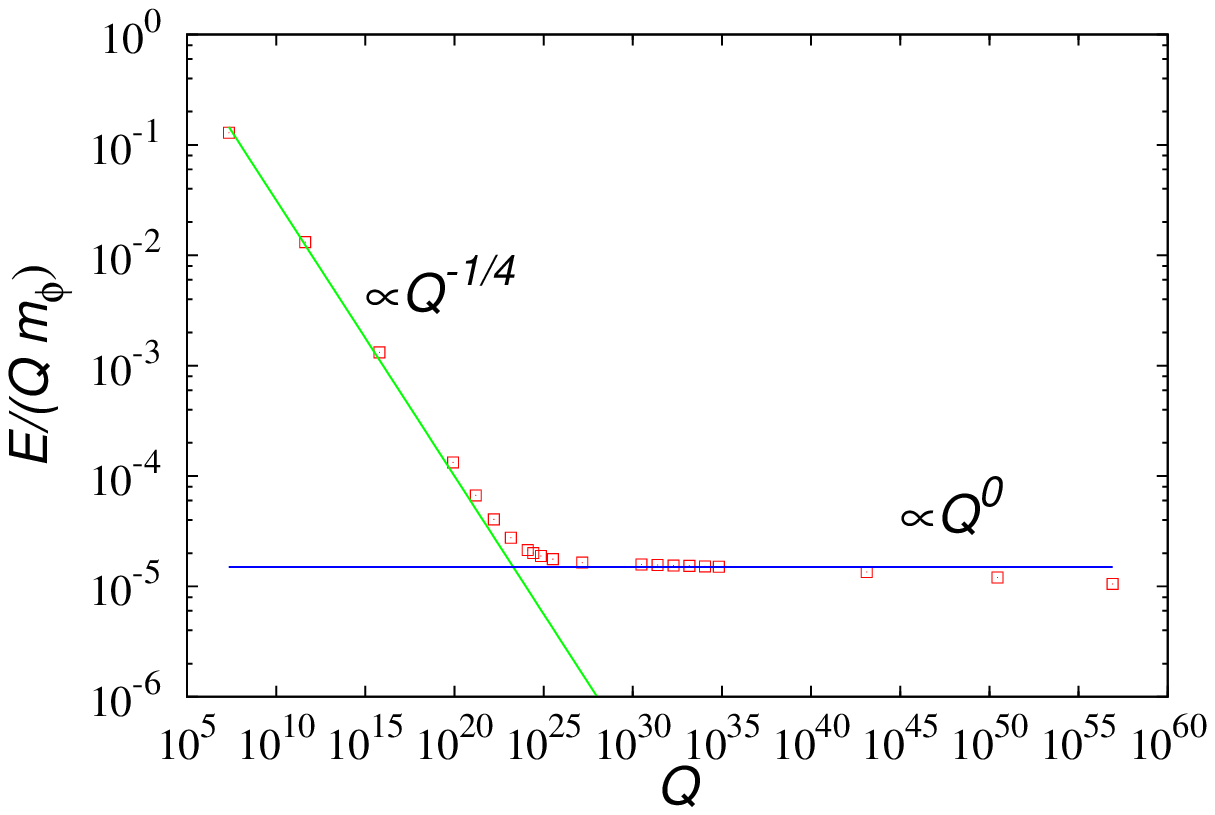}
\end{tabular}
\caption{Energy, size, field value at the center, and energy per charge of the $Q$ balls. 
Green and blue lines show the $Q$-dependence estimated analytically for the 
gauge-type and new-type $Q$ balls, respectively.}
\label{new-para}
\end{figure*}

The potential is written as \cite{New}
\begin{equation}
\label{pot-new}
V(\phi) = m_\phi^4\log\left( 1+\frac{|\phi|^2}{m_\phi^2}\right)
+m_{3/2}^2 |\phi|^2 \left( 1+K\log\frac{|\phi|^2}{M_*^2}\right),
\end{equation}
where the first (second) term comes from the gauge- (gravity-)mediation effect.
Here $m_\phi$ is the scalar mass in the vacuum, $m_{3/2}$ the gravitino mass,
$K<0$ the coefficient of the one-loop effect, and $M_*$  a renormalization scale. When the
each term of the potential dominates, the properties of each type of the $Q$ ball
are well known. For the gauge-type $Q$ ball, the energy $E$, the size $R$, 
the rotation speed $\omega$, and 
the field value at the center $\varphi_c$ are given by \cite{Dvali,KuSh,KK3}
\begin{eqnarray}
& & E \sim m_\phi Q^{3/4}, \nonumber \\
& & R \sim \omega^{-1} \sim m_\phi^{-1} Q^{1/4}, \nonumber \\ 
& & \varphi_c \sim m_\phi Q^{1/4}, 
\label{prop-gauge}
\end{eqnarray}
while for the new-type $Q$ ball \cite{New,EnMc},
\begin{eqnarray}
& & E \sim m_{3/2} Q, \nonumber \\
& & R \sim |K|^{-1/2} m_{3/2}^{-1}, \nonumber \\
& & \omega \sim m_{3/2}, \nonumber \\
& & \varphi_c \sim m_{3/2} Q^{1/2}.
\label{prop-new}
\end{eqnarray}
To look also for the transition region, we solve numerically Eq.(\ref{eom}) for
the potential (\ref{pot-new}). In Fig.~\ref{new-para}, the energy, size, the field value, 
and energy per charge as a function of the charge are shown.
In these figures, we set $m_{3/2}/m_\phi=10^{-5}$, $K=-0.01$, $M_*/m_\phi=10^{10}$.
One can see the features of both gauge-type and new-type $Q$ balls, {\it i.e.}, 
Eqs.(\ref{prop-gauge}) and (\ref{prop-new}), are reproduced in Fig.~\ref{new-para}.
Moreover, those parameters are smoothly connected in the transition region, as it should be. 
In particular, the energy per charge $E/Q$ always decreases as the charge $Q$ increases, 
which implies that the larger $Q$ ball is energetically favored. This shows that the new-type $Q$ ball
is stable against disintegration into smaller gauge-type $Q$ balls, so that it can be the dark 
matter of the universe, contrary to the claim in Ref.~\cite{Shoemaker}.

\section{Thin-wall-type $Q$ balls}
\begin{figure}[ht]
\includegraphics[width=85mm]{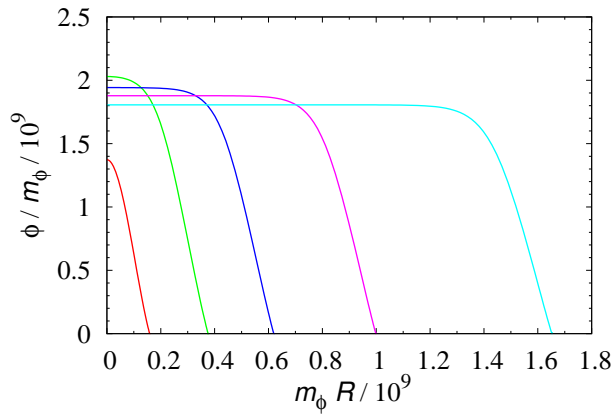}
\caption{Profile of the $Q$ balls for $\omega/m_\phi=2 \times 10^{-8}$, $1 \times 10^{-8}$,
$8 \times 10^{-9}$, $7 \times 10^{-9}$, and $6 \times 10^{-9}$ from the left to the right.
They correspond to $Q=9.4\times10^{34}$, $2.5\times 10^{36}$, $1.3\times 10^{37}$,
$6.0\times 10^{37}$, and $2.6\times 10^{38}$, respectively.}
\label{gaugeNR-prof}
\end{figure}

\begin{figure*}[ht]
\begin{tabular}{cc}
\includegraphics[width=85mm]{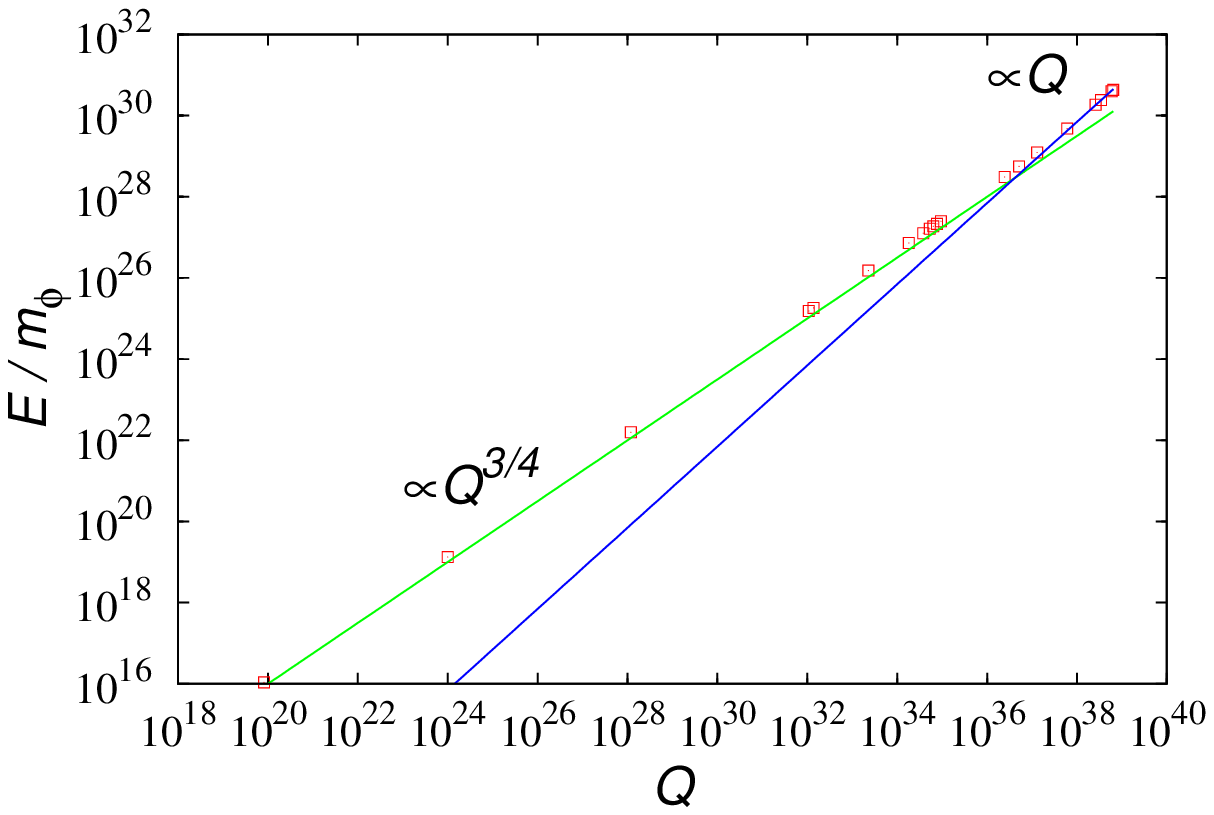} & 
\includegraphics[width=85mm]{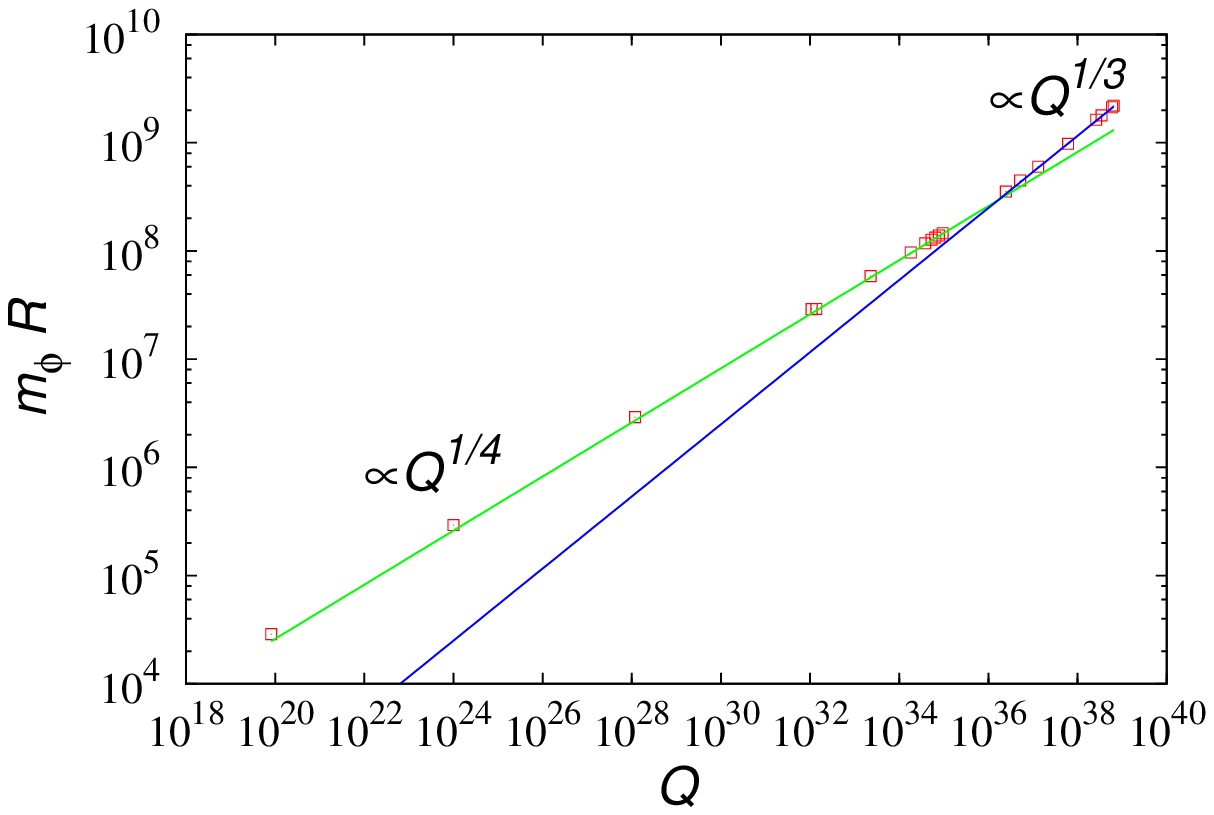} \\
\includegraphics[width=85mm]{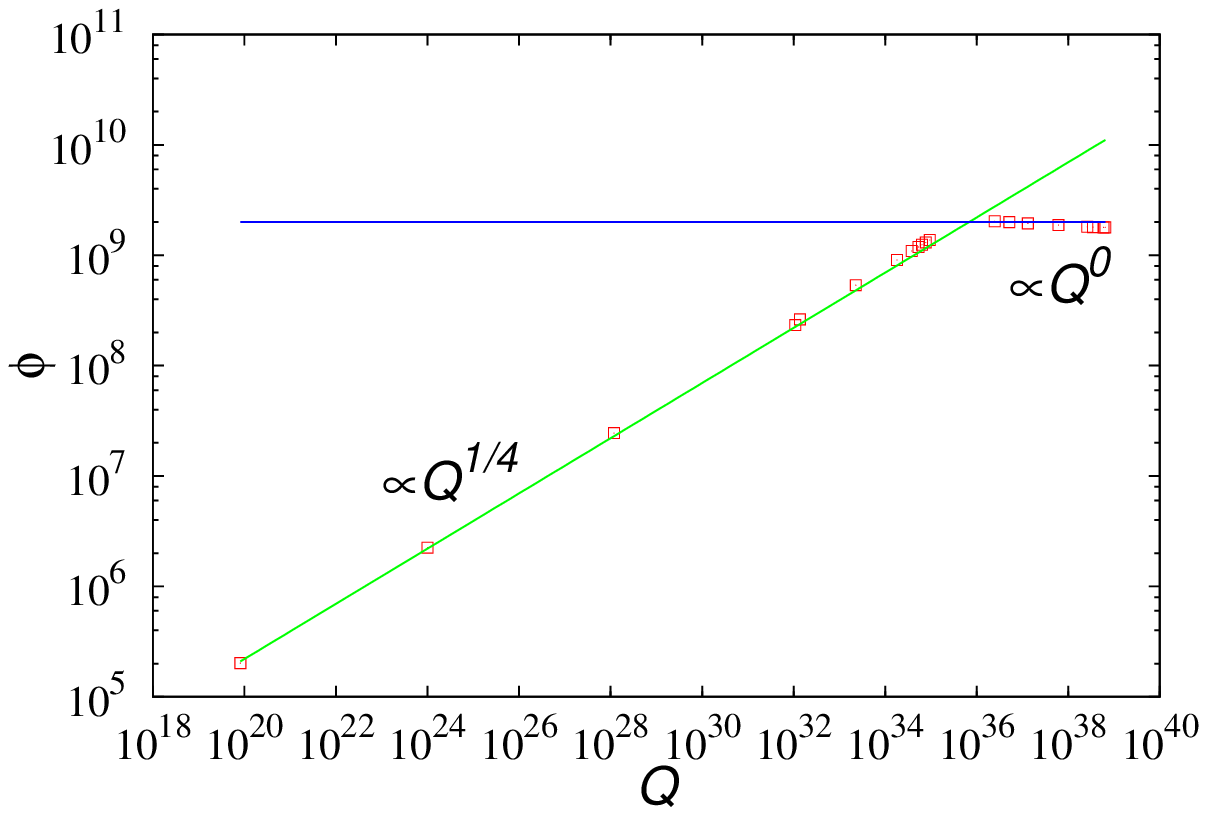} &
\includegraphics[width=85mm]{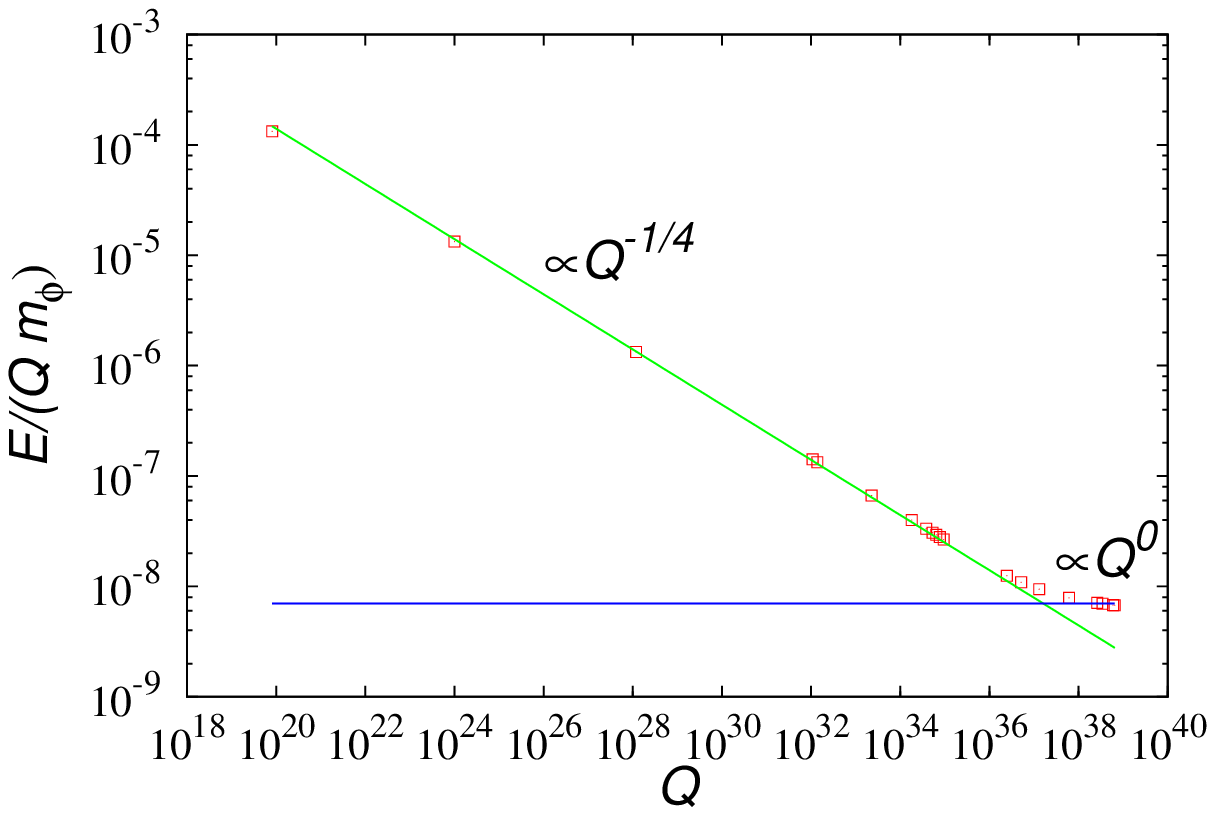}
\end{tabular}
\caption{Energy, size, field value at the center, and energy per charge of the $Q$ balls.
Green and blue lines show the $Q$-dependence estimated analytically for the 
gauge-type and thin-wall-type $Q$ balls, respectively.}
\label{gaugeNR-para}
\end{figure*}

Now let us consider the following potential,
\begin{equation}
\label{pot-gaugeNR}
V=m_\phi^4\log\left(1+\frac{|\phi|^2}{m_\phi^2}\right)+\frac{\lambda^2|\phi|^{2(n-1)}}{M_P^{2(n-3)}}.
\end{equation}
Although the thin-wall-type $Q$ ball is not created in the early universe, it could be formed 
through charge accumulation from the gauge-type $Q$ ball when the latter is swallowed by
the neutron star \cite{NS2}. The properties of the thin-wall-type $Q$ ball are also well known as 
\cite{Coleman,NS2}
\begin{eqnarray}
& & E \sim \mu Q, \nonumber \\
& & R \sim \left( \frac{\mu Q}{m_\phi^4}\right)^{1/3}, \nonumber \\
& & \omega \sim \mu \sim \frac{m_\phi^2}{\varphi_c}, \nonumber \\
& & \varphi_c \sim \left(\frac{m_\phi^2M_P^{n-3}}{\lambda}\right)^{1/(n-1)}.
\label{prop-thin}
\end{eqnarray}
To see the transition region as well, we solve numerically Eq.(\ref{eom}) for the potential
(\ref{pot-gaugeNR}). In Fig.~\ref{gaugeNR-prof}, we show the profile of the $Q$ balls, where
one can see the growth of the thin-wall-type $Q$ ball as well as the deformation from the 
gauge-type $Q$ balls as the charge is accumulating.
$Q$-ball properties are shown in Fig.~\ref{gaugeNR-para}. They coincide to analytical estimates
(\ref{prop-gauge}) and (\ref{prop-thin}) for the gauge-type and thin-wall-type $Q$ balls, respectively,
and they are smoothly connected in between.

\section{Conclusions}
We have revisited the new-type of the $Q$ ball, and  clarified its properties. In particular, we have
focused on the energy per charge $E/Q$ as a function of the charge $Q$, and reconfirmed the
stability of the new-type Q ball to be the dark matter of the universe. This corrects some misunderstandings in Ref.~\cite{Shoemaker}. In addition, we have investigated 
the feature of the $Q$ ball, transiting from the gauge-type to thin-wall-type $Q$ balls, 
which was considered in the context of the $Q$-ball capture by the neutron star in Ref.~\cite{NS2}.
Since the energy per charge $E/Q$ decreases as the charge $Q$ increases through the transition
region, we confirm the `metamorphosis' of the gauge-type to thin-wall-type $Q$ balls.

\section*{Acknowledgments}
This work is supported by Grant-in-Aid for Scientific research from the Ministry of 
Education, Science, Sports, and Culture (MEXT), Japan, under Contract No. 14102004 (M.K.),
and also by World Premier International Research Center Initiative, MEXT, Japan.



\end{document}